\documentclass[english]{article}
\usepackage{textcomp}
\usepackage{amsmath}
\usepackage{amssymb}
\usepackage{graphicx}
\usepackage[left=1.75cm,top=1.5cm,right=1.75cm,bottom=2.5cm]{geometry}
\begin{document}
\begin{center}
\Large{\bf{$\Lambda$-CDM  type Heckmann - Suchuking model and union 2.1 compilation}}\\
\vspace{4mm}
\normalsize{G. K. Goswami$^1$, R. N. Dewangan$^2$ \& Anil Kumar Yadav$^3$ }\\
\vspace{2mm}
\normalsize{$^{1, 2} $Department of Mathematics, Kalyan P. G. College, Bhilai - 490006, India}\\
\vspace{1mm}
\normalsize{Email: gk.goswam9@gmail.com}\\
\vspace{2mm}
\vspace{2mm}
\normalsize{$^3$ Department of Physics, United College of Engineering \& Research,\\ Greater Noida - 201306, India}\\
\vspace{1mm}
\normalsize{Email: abanilyadav@yahoo.co.in}\\
\end{center}
\vspace{2mm}
\begin{abstract}
In this paper, we have investigated $\Lambda$-CDM   type cosmological model in
Heckmann-Schucking space-time, by using 287  high red shift ($ .3 \leq z \leq 1.4$ ) SN Ia
data of observed absolute magnitude along with their
possible error from Union 2.1 compilation. We have used $\chi^{2}$ test
to compare  Union 2.1 compilation observed data and
corresponding theoretical values of apparent magnitude $(m)$. It is found that the best fit value
for  $(\Omega_{m})_0$, $(\Omega_{\Lambda})_0$ and $(\Omega_{\sigma})_0$ are $0.2940$, $0.7058$ and $0.0002$ respectively
and the derived model represents the features of accelerating universe which is consistent with recent
astrophysical observations.\\
\end{abstract}

\section{Introduction}
Wilkinson Microwave Anisotropy Probe (WMAP)[25] and
Hubble Key Project (HKP)[9] explored that our universe is nearly flat.
This has given concept of two component density parameters $\Omega_{m}$ and $\Omega_{\Lambda}$ ,
which are related through
\begin{equation}\label{eq1}
\Omega_{\Lambda}+\Omega_{m}=1.
\end{equation}
Equation (1) is obtained by solving Einstein's Field
Equations with cosmological constant for  FRW cosmological model,
which represent a spatially homogeneous and isotropic accelerating expanding flat universe. One can see the details of
$\Lambda$-CDM model in Refs. [1-18].\\
Luminosity distance $(D_{L})$ in $\Lambda$- CDM model is as follows:
\begin{equation}
\label{eq2}
D_{L}=\frac{cz}{H_{0}},
\end{equation}
and
\begin{equation}\label{eq3}
D_{L}=\frac{c(1+z)}{H_{0}}\int_{0}^{z}\frac{dz}{\sqrt{[(\Omega_{m})_{0}(1+z)^{3}+(\Omega_{\Lambda})_{0}]}}.
\end{equation}
The luminosity distance $(D_{L})$ is associated with absolute and
apparent magnitudes by the following equation
\begin{equation}\label{eq4}
m-M=5log_{10}(\frac{D_{L}}{Mpc})+25.
\end{equation}
To get the absolute magnitude $M$ of a supernova, we consider a supernova 1992P at
low-redshift z = 0.026 with m = 16.08.\\
Equations (\ref{eq2}) and (\ref{eq4}) read as
\begin{center}
\begin{equation}
\label{eq5}
M=16.08-25+5log_{10}(H_{0}/.026c).
\end{equation}
\par\end{center}
These equations (\ref{eq2})-(\ref{eq5}) produce the following expression for absolute
magnitude $m$.
\begin{equation}
m
=16.08+5log_{10}(\frac{(1+z)}{.026}\int_{0}^{z}\frac{dz}{\sqrt{[(\Omega_{m})_{0}(1+z)^{3}+(\Omega_{\Lambda})_{0}]}})).
\end{equation}
In the last decade of $20^{th}$ century, Riess et al. [20] and Perlmutter et al. [18]
 found that the present
values of $\Omega_{m}$ and $\Omega_{\Lambda}$ are nearly $0.29$ and
$0.71$ respectively. Perlmutter et al. had used only 60 SN Ia low red shift data set while in the
present analysis, we have used 287 high red shift data out of 500 SN Ia data set as reported in ref. [26].
The recent SN Ia observations, BOSS, WMAP and Plank result for CMB anisotropy [12] give more precise
value of cosmological parameters. After publication of WMAP data, we notice that today, there is considerable evidence in support of
anisotropic model of universe. On the theoretical front, Misner [16] has investigated
an anisotropic phase of universe, which turns into isotropic one. The authors of ref. [17]
have investigated the accelerating model of universe with anisotropic EOS parameter and have also shown that
the present SN Ia data permits large anisotropy. Recently DE models with variable EOS parameter in anisotropic
space-time have been studied by Yadav and Yadav [27], Yadav et al [28,29],
Akarsu and Kilinc [3], Yadav [30], Saha and Yadav [22] and Pradhan [19].\\

In Ref. [10], we have presented a $\Lambda$-CDM   type cosmological model in spatially
homogeneous and anisotropic Heckmann-Schucking space-time given by
\begin{equation}\label{metric}
 ds^{2}= c^{2}dt^{2}- A^{2}dx^{2}-B^{2}dy^{2}-C^{2}dz^{2},
\end{equation}
Where $A(t)$, $B(t)$ and $C(t)$ are scale factors along $x$, $y$ and $z$ axes. In the literature, metric (\ref{metric})
is also named as Bianchi type I [31]. \\
We consider energy momentum tensor for a perfect fluid  i.e.
\begin{equation}\label{emt}
T_{ij}=(p+\rho)u_{i}u_{j}-pg_{ij},
\end{equation}
Where $g_{ij}u^{i}u^{j}=1$ and $u^{i}$ is the 4-velocity vector.\\
In co-moving co-ordinates
\begin{equation}
u^{\alpha}=0,~~~~~~~~~\alpha=1,2, 3.
\end{equation}
The Einstein field equations are
\begin{equation}
\label{efe}
R_{ij}-\frac{1}{2}Rg_{ij} + \Lambda g_{ij}= -\frac{8\pi
G}{c^{4}}T_{ij}.
\end{equation}
Choosing co-moving coordinates, the field equations (\ref{efe}), for the line element (\ref{metric}), read as
\begin{equation}
\label{fe1}
\frac{\ddot{B}}{B}+\frac{\ddot{C}}{C}+\frac{\dot{B}\dot{C}}{BC}=-\frac{8\pi G}{c^{2}} p+\Lambda
c^{2},
\end{equation}
\begin{equation}\label{fe2}
\frac{\ddot{A}}{A}+\frac{\ddot{C}}{C}+\frac{\dot{A}\dot{C}}{AC}=- \frac{8\pi G}{c^{2}}p+\Lambda
c^{2},
\end{equation}
\begin{equation}\label{fe3}
\frac{\ddot{A}}{A}+\frac{\ddot{B}}{B}+\frac{\dot{A}\dot{B}}{AB}=- \frac{8\pi G}{c^{2}}p+\Lambda
c^{2},
\end{equation}
\begin{equation}\label{fe4}
 \frac{\dot{A}\dot{B}}{AB}+\frac{\dot{B}\dot{C}}{BC}+\frac{\dot{C}\dot{A}}{AC}= \frac{8\pi G}{c^{2}}\rho+\Lambda
 c^{2}.
\end{equation}
Solving equations (\ref{fe1}) - (\ref{fe3}) by approach given in ref. [10], one can obtain the following
relation for scale factors
\begin{equation}
 \label{sf1}
B = AD(t),
\end{equation}
\begin{equation}
\label{sf2}
C = AD(t)^{-1},
\end{equation}
\begin{equation}
 \label{sf3}
D(t) = exp\left[\int\frac{K}{A^{3}}\right],
\end{equation}
Where $K$ is the constant of integration.\\
In view of equations (\ref{sf1}) - (\ref{sf3}), equations (\ref{fe1}) - (\ref{fe4}), read as
\begin{equation}\label{fe5}
2\frac{\ddot{A}}{A}+\frac{\dot{A}^{2}}{A^{2}}=-\frac{8\pi
G}{3c^{2}}\Bigl(p-\frac{\Lambda c^{4}}{8\pi
G}+\frac{K^{2}c^{2}}{8\pi GA^{6}}\Bigr).
\end{equation}
\begin{equation}\label{fe6}
H^{2}=\frac{\dot{A}^{2}}{A^{2}}=\frac{8\pi
G}{3c^{2}}\Bigl(\rho+\frac{\Lambda c^{4}}{8\pi
G}+\frac{K^{2}c^{2}}{8\pi GA^{6}}\Bigr).
\end{equation}
Now, we assume that the cosmological constant $\Lambda$ and the
term due to anisotropy also act like energies with densities and
pressures as
$$\rho_{\Lambda}=\frac{\Lambda c^{4}}{8\pi G}\hspace{.5in}
\rho_{\sigma}=\frac{K^{2}c^{2}}{8\pi GA^{6}}$$
\begin{equation}
p_{\Lambda}=-\frac{\Lambda c^{4}}{8\pi G}\hspace{.5in}
p_{\sigma}=\frac{K^{2}c^{2}}{8\pi GA^{6}}.
\end{equation}
 It can be easily verified that energy conservation law holds
 separately for $\rho_{\Lambda}$ and $\rho_\sigma$ i.e.

 $$\dot{\rho_{\Lambda}}+3H(p_\Lambda+\rho_\Lambda)=0,
 $$
$$\dot{\rho_{\sigma}}+3H(p_\sigma+\rho_\sigma)=0.
 $$\\
 The equations of state for matter, $ \sigma$ and $\Lambda$ energies are
 read as
 \begin{equation}
 p_m=\omega_m\rho_m,
\end{equation}
where $\omega_m = 0$ for matter in form of dust,
$\omega_m=\frac{1}{3}$ for matter in form of radiation. There are
certain more values of $\omega_m$ for matter in different forms
during the course of evolution of the universe.
 \begin{equation}
 p_{\Lambda}=\omega_{\Lambda}\rho_{\Lambda},
\end{equation}
 Since
 $$p_{\Lambda}+ \rho_{\Lambda}=0,$$
  Therefore
  $$\omega_{\Lambda}=-1,$$
  Similarly
$$p_{\sigma}=\rho_{\sigma},$$
So,
$$\omega_{\sigma}=1,$$
where $\omega_{m}$, $\omega_{\Lambda}$ and $\omega_{\sigma}$ are equation of state parameters of matter,
$\Lambda$ and $\sigma$.\\
In the literature, $\Lambda$-CDM   model is described by FLRW metric. In the derived model, equations (\ref{fe5}) and
(\ref{fe6}) represent the field equation of FLRW metric that is why the derived model is $\Lambda$-CDM   type. For flat
$\Lambda$-CDM   model, we have
\begin{equation}
\label{eq1}
\Omega_{\Lambda}+\Omega_{m}+ \Omega_{\sigma} = 1,
\end{equation}
where $
 \Omega_{m}=\frac{\rho_{m}}{\rho_{c}}
=\frac{(\Omega_m)_0H^2_0(1+z)^3}{H^2}$, $
\Omega_{\Lambda}=\frac{\rho_{\Lambda}}{\rho_{c}}
=\frac{(\Omega_\Lambda)_0H^2_0}{H^2}$ and
$\Omega_{\sigma}=\frac{\rho_{\sigma}}{\rho_{c}}
=\frac{(\Omega_\sigma)_0H^2_0(1+z)^6}{H^2}.$\\
\\
$\Omega_{\sigma}$ is the anisotropic energy density which is taken very small for present analysis. The recent
observation of CMB support the existence of anisotropy in early phase of evolution of universe. Therefore, it
make sense to consider the model of universe with an anisotropic background in the presence of cosmological term $\Lambda$.\\

The expression for luminosity distance $(D_{L})$ and apparent magnitude $(m)$
 are as follows
\begin{equation}
D_{L}=\frac{c(1+z)}{H_{0}}\int^z_0\frac{dz}{\sqrt{[(\Omega_m)_0(1+z)^3+(\Omega_\sigma)_0(1+z)^6+(\Omega_\Lambda)_0]}}.
\end{equation}

\begin{equation}
 m =16.08 +5log_{10}(\frac{(1+z)}{.026}\int_{0}^{z}\frac{dz}
{\sqrt{[(\Omega_m)_0(1+z)^3+(\Omega_\sigma)_0(1+z)^6+(\Omega_\Lambda)_0]}}).
\end{equation}
The purpose of the present work is to compute the $\Omega_{\Lambda}$ and
the $\Omega_{m}$ of the universe in light of Union $2.1$ compilation in anisotropic space-time which is
different from our previous work [10]. In our previous work [10], we used old SN Ia data
to compute the physical parameters at present epoch. Here the anisotropic energy density $(\Omega_{\sigma})$ is taken
to be very small i. e.
$\Omega_{\sigma}$ = 0.0002. In order to obtain $\Omega_{\Lambda}$
and $\Omega_{\sigma}$,  we have considered high red-shift SN Ia supernova data of observed $m$
along with their possible error from Union $2.1$
compilation and have obtained corresponding theoretical values of
$m$ for various $\Omega_m$, ranging in between $0$ and $1$. The paper is organized as follows.
In section 2, we have estimated the
values of $\Omega_{\Lambda}$ and $\Omega_{m}$. Section 3 deals with the matter and dark energy densities and
estimation of present age of universe. Finally the results are displayed in section 4.\\

\section{Estimation of $(\Omega_{m})_{0}$ and $(\Omega_{\Lambda})_{0}$ from 287 SN Ia data set}
For the sake of comparison of theoretical value of $\Omega_{m}$ with observational values, we compute the
$\chi^{2}$ value as following:
\begin{center}
{\large{$\chi_{SN}^{2}=A-\frac{B^{2}}{C}+log_{10}(\frac{C}{2\pi})$}},
\par\end{center}{\large \par},

{\large{$ $}}{\large \par}
where
\begin{center}
{\large{$A =\overset{287} {\underset{i=1}{\sum}}\frac{\left[\left(m\right)_{ob}-\left(m\right)_{th}\right]^{2}}
{\sigma_{i}^{2}}$}},
\par\end{center}
{\large \par}

\begin{center}
{\large{$B =\overset{287} {\underset{i=1}{\sum}}\frac{\left[\left(m\right)_{ob}-\left(m\right)_{th}\right]}
{\sigma_{i}^{2}}$}},
\par\end{center}{\large \par}

{\large{$ $}}{\large \par}
and
\begin{center}
{\large{$C =\overset{287} {\underset{i=1}{\sum}}\frac{1}{\sigma_{i}^{2}}$}}.
\par\end{center}{\large \par}

Here the summations are taken over data sets of observed and theoretical values of apparent magnitudes of 287 supernovae. \\

Based on the above expressions, we obtain
the following Table-1 which describes the various values of  $\chi^{2}$
against values of $\Omega_{m}$ ranging  between $0$ to $1$
\begin{center}
\begin{tabular}{|c|c|c|}
\hline $\Omega_{m}$ & $\chi_{SN}^{2}$ &
$\chi_{SN}^{2}/287$\tabularnewline \hline 0.0000  &
5522.1000   &   19.24076655 \tabularnewline \hline 0.1000  &
50002.1000  &   174.22334495    \tabularnewline \hline 0.2000  &
4833.7000   &   16.84216028 \tabularnewline \hline 0.2800  &
4799.8000   &   16.72404181 \tabularnewline \hline 0.2900  &
4799.2000   &   16.72195122 \tabularnewline \hline 0.2940  &
4799.3000   &   16.72229965 \tabularnewline \hline 0.9998  &
4799.2000   &   16.72195122 \tabularnewline \hline 0.2960  &
4799.3000   &   16.72229965 \tabularnewline \hline 0.9996  &
5276.7000   &   18.38571429 \tabularnewline  \hline
\end{tabular}
\par\end{center}
\begin{center}
\textbf{Table: 1}
\end{center}
$\vphantom{}$

 From Table-1, we find that for minimum value of $\chi^2$,  the best fit present values of $\Omega_{m}$ and
  $\Omega_{\Lambda}$  are  $0.2940$ and $0.7058$ respectively.\\

In order to compare the various theoretical value of luminosity distance
$D_L$ corresponding to different values of $\Omega_{m}$ with observational values,  we
compute again $\chi^{2}$ values as following:
\begin{center}
{\large{$\chi_{SN}^{2}=A-\frac{B^{2}}{C}+log_{10}(\frac{C}{2\pi})$}},
\par\end{center}{\large \par},

{\large{$ $}}{\large \par}
where
\begin{center}
{\large{$A=\overset{287}{\underset{i=1}{\sum}}\frac{\left[\left(D_L\right)_{ob}-\left(D_L\right)_{th}\right]^{2}}
{\sigma_{i}^{2}}$}},
\par\end{center}
{\large \par}

\begin{center}
{\large{$B=\overset{287}{\underset{i=1}{\sum}}\frac{\left[\left(D_L\right)_{ob}-\left(D_L\right)_{th}\right]}
{\sigma_{i}^{2}}$}},
\par\end{center}{\large \par}

{\large{$ $}}{\large \par}
and
\begin{center}
{\large{$C=\overset{287}{\underset{i=1}{\sum}}\frac{1}{\sigma_{i}^{2}}$}}.
\par\end{center}{\large \par}
$\vphantom{}$
$\vphantom{}$\\

Here the summations are over data sets of observed and theoretical values of luminosity distances of 287 supernovae. \\

Based on the above expressions , we obtain
the following Table-2 which describes the various values of  $\chi^{2}$
against values of $\Omega_{m}$ ranging  between $0$ to $1$
\begin{center}
\begin{tabular}{|c|c|c|}
\hline $\Omega_{m}$ & $\chi_{SN}^{2}$ &
$\chi_{SN}^{2}/287$\tabularnewline \hline  0.0000  &
225.5970    &   0.78605226  \tabularnewline \hline 0.1000  &
204.7956    &   0.71357352  \tabularnewline \hline 0.2000  &
198.0609    &   0.69010767  \tabularnewline \hline 0.2800  &
196.7042    &   0.68538049  \tabularnewline \hline 0.2900  &
196.6806    &   0.68529826  \tabularnewline \hline 0.2910  &
196.6797    &   0.68529512  \tabularnewline \hline 0.2920  &
196.9790    &   0.68633798  \tabularnewline \hline 0.2930  &
196.6785    &   0.68529094  \tabularnewline \hline 0.2940  &
196.6783    &   0.68529024  \tabularnewline \hline 0.2960  &
196.6786    &   0.68529129  \tabularnewline \hline 0.2970  &
196.6791    &   0.68529303  \tabularnewline \hline 0.3000  &
196.6821    &   0.68530348  \tabularnewline \hline 0.9998  &
215.6214    &   0.75129408  \tabularnewline \hline
\end{tabular}
\par\end{center}
\begin{center}
\textbf{Table: 2}
\end{center}

From Table-2, we find that for minimum value of $\chi^2$,  the best fit present values of $\Omega_{m}$ and
  $\Omega_{\Lambda}$ are presented again as $(\Omega_{m})_0$ = $0.2940$ and  $(\Omega_{\Lambda})_0$ = $0.7058$ respectively.\\

\section{Some physical parameters of the universe}
\subsection{Density of the universe and Hubble's constant}
$ $
The energy density at present is given by
\begin{equation}
\label{eq7}
(\rho_{i})_{0}=\frac{3c^{2}H_{0}^{2}}{8\pi G}(\Omega_{i})_{0},
\end{equation}
where i stands for different types of energies such as matter energy,
dark energy etc.\\
Taking, $(\Omega_{m})_{0}$ = 0.2940, $H_{0}=72km/sec./Mpc$.\\
The current value of dust energy $(\rho_{m})_{0}$ for flat
universe is given by
\begin{equation}
\label{eq8} (\rho_{m})_{0}= 0.5527 h^2_0\times10^{-29}gm/cm^{3}.
\end{equation}
The current value of dark energy $(\rho_{\Lambda})_{0}$ read as
\begin{equation}
\label{eq9} (\rho_{\Lambda})_{0}=\frac{3c^{2}H_{0}^{2}}{8\pi
G}(\Omega_{\Lambda})_{0}=  1.3269 h^2_0\times10^{-29}gm/cm^{3},
\end{equation}
Where $(\Omega_{\Lambda})_{0}$= 0.7058.\\

The expression for Hubble parameter is given by
\begin{equation}
\label{eq10}
H^{2}=H_{0}^{2}[(\Omega_m)_0(1+z)^3+(\Omega_\sigma)_0(1+z)^6+(\Omega_\Lambda)_0]\,
\end{equation}
and
\begin{equation}
\label{eq11}
H^{2}=H_{0}^{2}[(\Omega_{m})_{0}\left(\frac{A_{0}}{A}\right)^{3}+(\Omega_\sigma)_0\left(\frac{A_{0}}{A}\right)^{6}+
(\Omega_{\Lambda})_{0}].
\end{equation}

\subsection{Age of the universe}
The present age of the universe is obtained as follows
\begin{equation}\label{eqt}
   t_{0}=\int^{t_0}_{0}dt=
\int^{\infty}_0\frac{dz}{H_0 (1+z)\sqrt{\bigl[(\Omega_m)_0(1+z)^3+(\Omega_\sigma)_0(1+z)^6+(\Omega_\Lambda)_0\bigr]}},
\end{equation}
\begin{equation}\label{eqt1}
t_{0}=\int^{t_0}_{0}dt=
\int^{\infty}_0\frac{dz}{H_0 (1+z)\sqrt{\bigl[(\Omega_m)_0(1+z)^3+(\Omega_\sigma)_0(1+z)^6+(\Omega_\Lambda)_0\bigr]}}.
\end{equation}
From equation (\ref{eqt1}), one can easily obtain $t_0\rightarrow 0.9388H_0^{-1}$ for high
redshift and $(\Omega_\Lambda)_0=0.7058$. This means that the present age of
the universe is 12.7534 Gyrs $\sim$13 Gyrs for $\Lambda$ dominated
universe. From WMAP data, the empirical value of present age of
universe is $13.73_{-.17}^{+.13}Gyrs$ which is closed to
present age of universe, estimated in the this paper.

\subsection{Deceleration parameter $q$:}
 The deceleration
parameter is given by
\begin{equation}
q = \frac{3}{2}\Bigl( \frac{(\Omega_m)_o (1+z)^3 +
2(\Omega_\sigma)_o (1+z)^6}{(\Omega_m)_o
(1+z)^3+(\Omega_\Lambda)_o + (\Omega_\sigma)_o
 (1+z)^6}\Bigr)-1.
\end{equation}
Since in the derived model, the best fit values of
$(\Omega_{m})_{0}$, $(\Omega_{\Lambda})_{0}$ and
$(\Omega_{\sigma})_{0}$ are 0.2940, 0.7058 and 0.0002 respectively
hence we compute the present value of deceleration parameter for
derived $\Lambda$-CDM universe by putting $z = 0$ in eq. (19). The
present value of DP comes to
\begin{equation}
 q_{0} = -0.5584.
\end{equation}
Also it is evident that the universe had entered in the accelerating phase
at $z\thicksim0.6805\backsimeq t\thicksim 0.4442 H^{-1}_0\thicksim 6.0337\times10^9 yrs$ in
the past before from now.\\
\section{Result and Discussion}
In the present work, $\Lambda$-CDM  type cosmological model
in Heckmann-Suchuking space-time has been investigated.
We summarize our work by presenting the following table which
displays the values of cosmological parameters at present.
\begin{center}
\begin{tabular}{|c|c|c|c|c|}
  \hline
  Cosmological Parameters & Values at Present \\
  \hline
  $(\Omega_\Lambda)_0$ & .7058 \\
   $(\Omega_m)_0$ & .2940\\
   $(\Omega_\sigma)_0$ & .0002 \\
  $(q)_0$& -0.5584\\
  $\small{(\rho_{m})_{0}}$ &$0.5527 h^2_0\times10^{-29}gm/cm^{3}$  \\
  $\small{(\rho_{\Lambda}})_{0}$ & $1.3269h^2_0\times10^{-29}gm/cm^{3}$\\
  $\small{Age~ of~ the~ universe}$ &$12.7534~Gyrs $\\
\hline
\end{tabular}
\end{center}
The figures 1 and 2 shows how the observed values of apparent magnitudes
and luminosity distances come close to the  theoretical graphs for
$(\Omega_\Lambda)_0 = 0.7058$, $(\Omega_m)_0 = 0.2940$ and
$(\Omega_\sigma)_0 = 0.0002$. Figures 3 and 4 shows the
dependence of Hubble's constant with
red shift and scale factors. Figure 5 shows that the time tends to a definite value for large
redshift which in turn determines the age of universe.
The various figures validates that the best fit value for energy parameters corresponding to matter and
dark energy are $0.2940$ and $0.7058$.
As a final comment, we note that the
present model represents the features of accelerating universe. The present value of DP is found $-0.5584$ which
is consistence with modern astrophysical observations.\\
\begin{center}
\includegraphics[width=9cm,height=9cm]{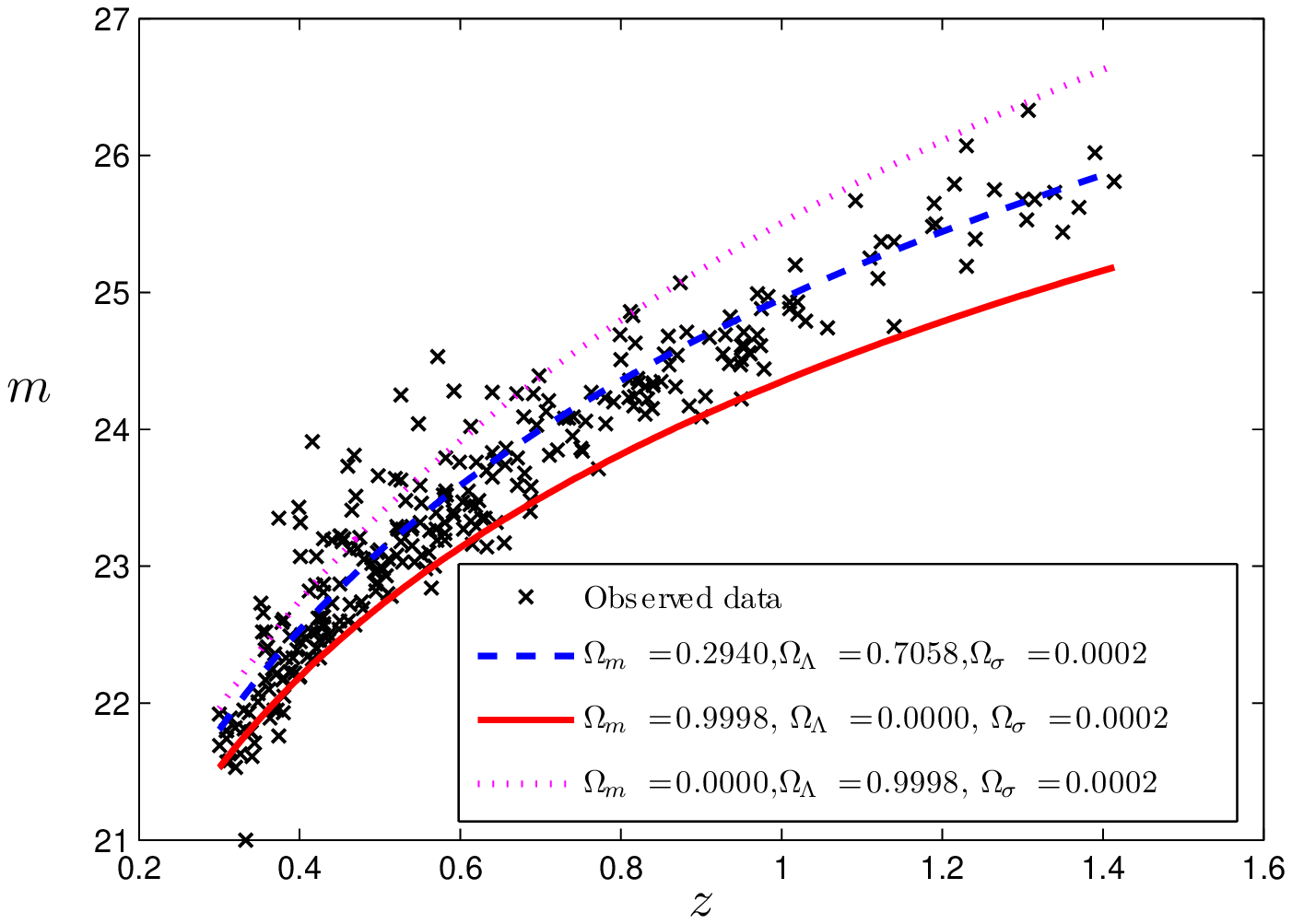}
\end{center}
\begin{center}
\textbf{Fig1:~Apparent magnitude $(m)$ versus redshift $(z)$}
\par\end{center}
\begin{center}
\includegraphics[width=11cm,height=11cm]{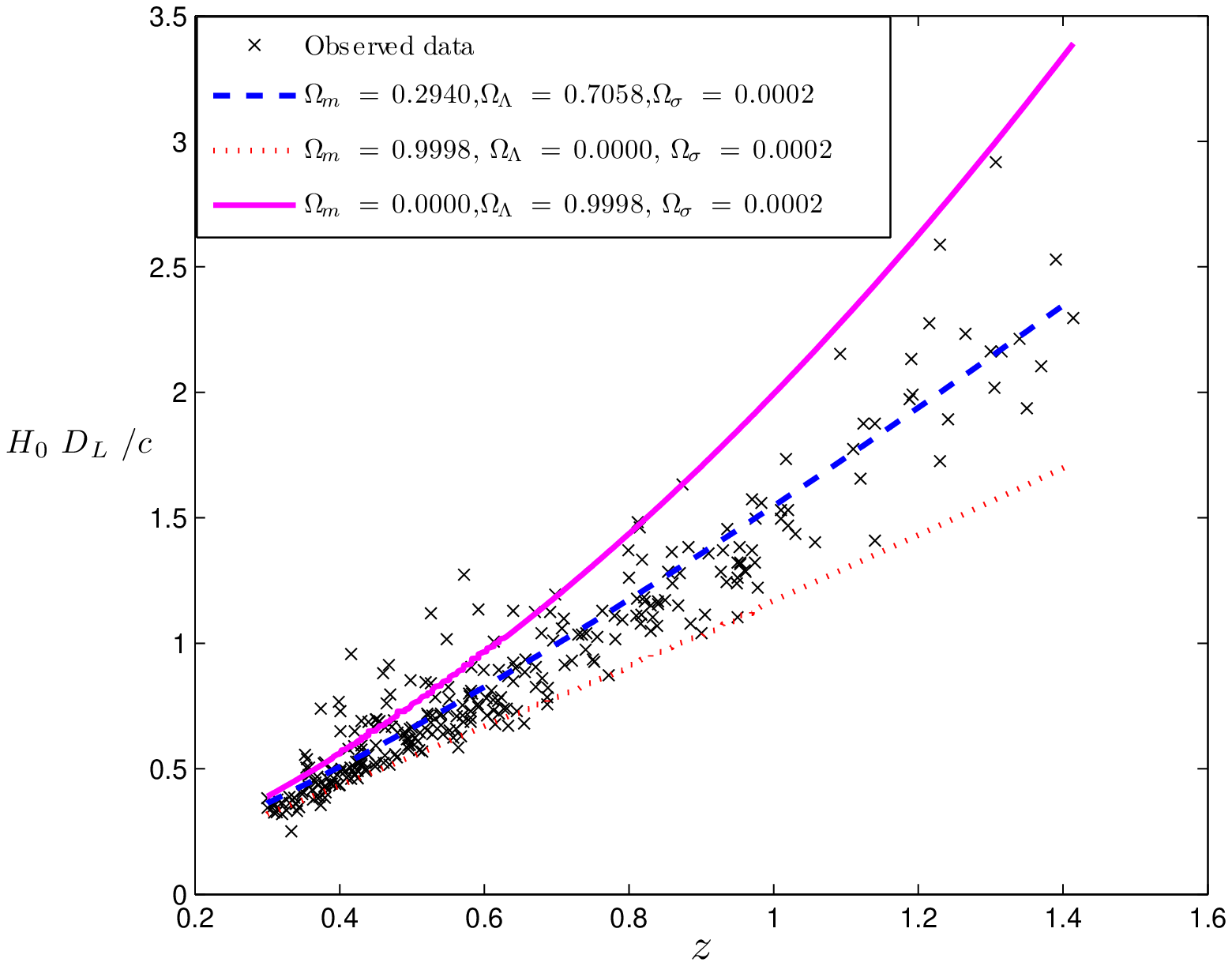}
\end{center}
\begin{center}
\textbf{Fig2:~ Luminosity distance $(D_{L})$ versus redshift $(z)$}
\par\end{center}
\begin{center}
\includegraphics[width=9cm,height=9cm]{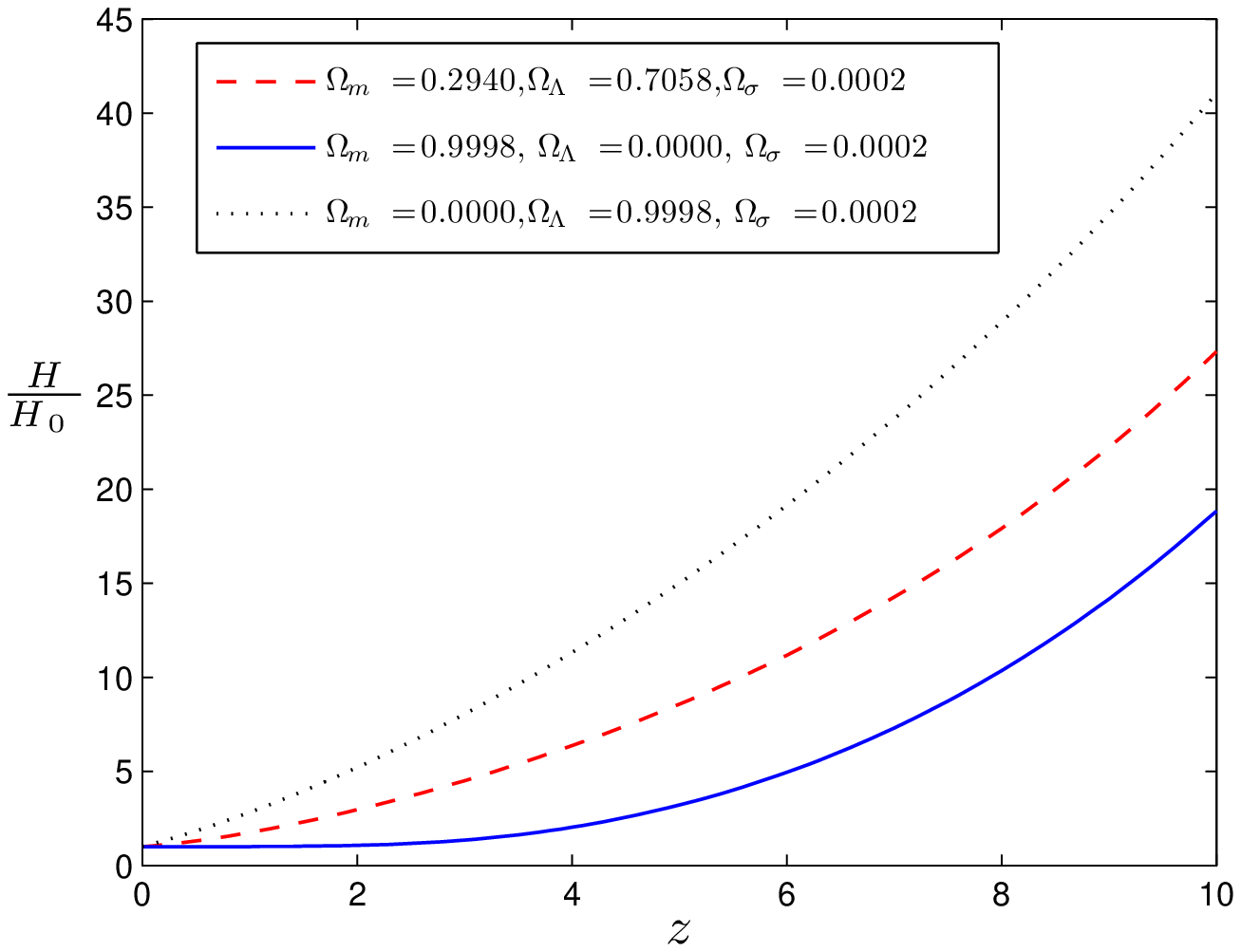}
\par\end{center}
\begin{center}
\textbf{Fig3:~ Hubble constant $\left(\frac{H}{H_{0}}\right)$
versus redshift $(z)$}
\par\end{center}
\begin{center}
\includegraphics[width=9cm,height=9cm]{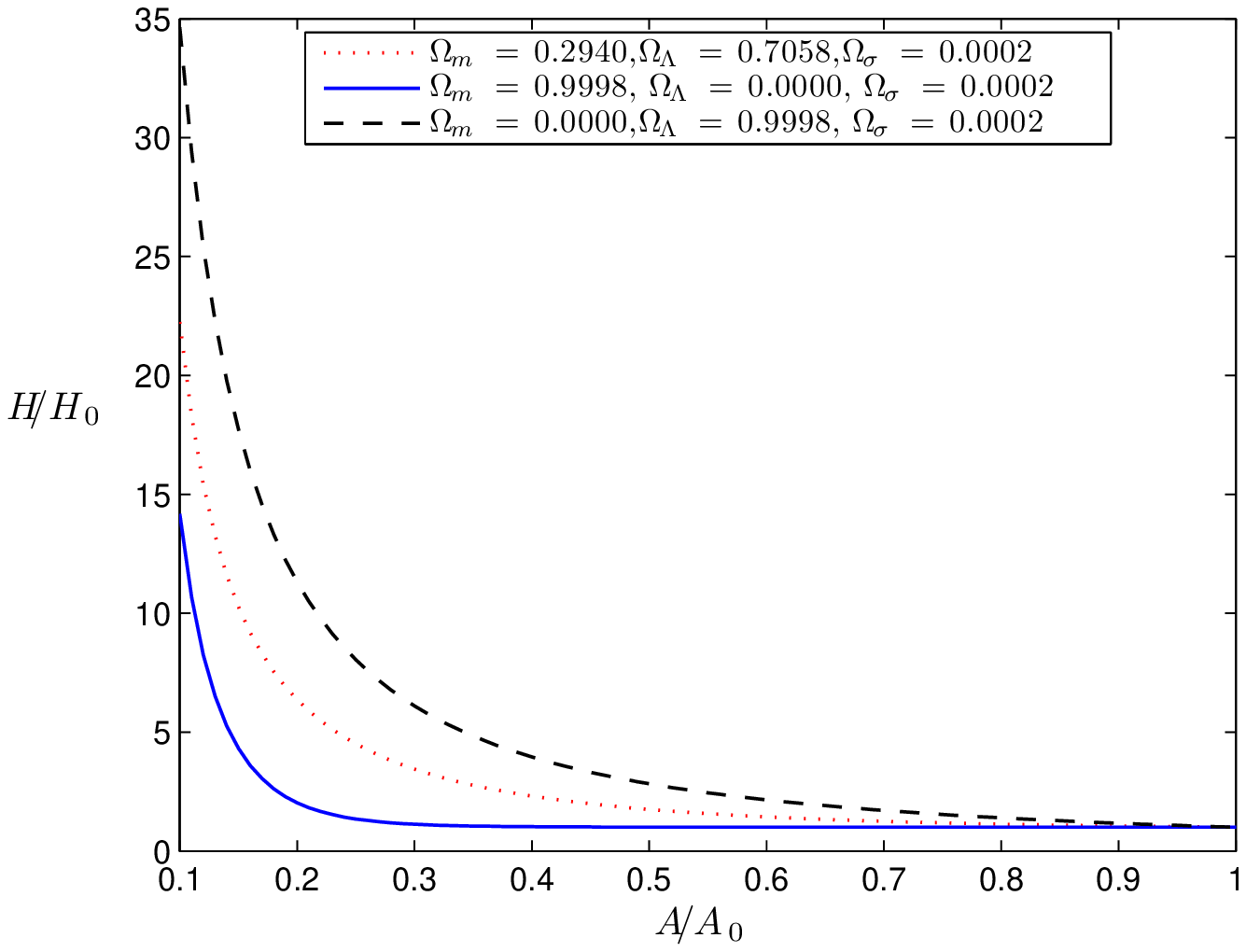}
\par\end{center}
\begin{center}
\textbf{Fig4:~ Hubble constant $\left(\frac{H}{H_{0}}\right)$
versus scale factor}
\par\end{center}
\begin{center}
\includegraphics[width=9cm,height=9cm]{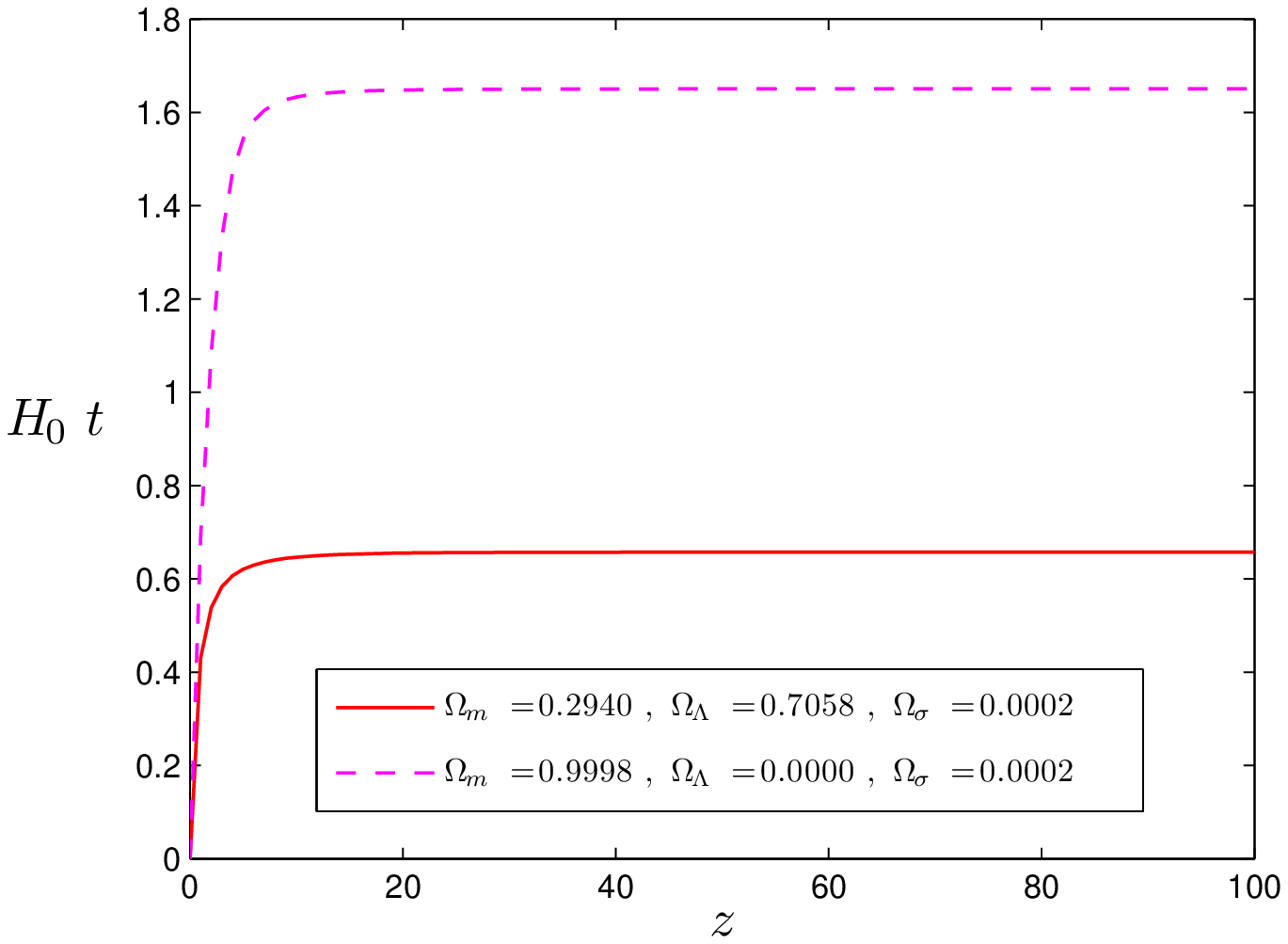}
\par\end{center}
\begin{center}
\textbf{Fig5:~ Time versus redshift $(z)$}
\par\end{center}
\section*{Acknowledgments}
The authors are grateful to the anonymous referee for valuable comments to improve the quality of manuscript. One of us G. K. G. is thankful to IUCAA, Pune, India for providing facility and support where part of this work was carried out during a visit.
This work is  supported by the CGCOST  Research Project 789/CGCOST/MRP/14.



\small
\begin{thebibliography}{99}
\bibitem{1} P. A. R. Ade et al.,`` The Second Planck Catalogue of Compact Sources,'' arXiv: 1303.5076v3.
\bibitem{2} U. Alam, V. Sahni, T D Saini and  A A Starobinsky, Month. Not. Roy. Astron. Soci. {\bf 344}, 1057 (2003).
\bibitem{3} O. Akarsu and C. B. Kilinc, Gen. Relativ. grav. {\bf 42}, 119 (2010).
\bibitem{4} R. Amanullah et al., Astrophys. J. {\bf 716}, 712 (2010)
\bibitem{5} P. Astier et al., Astron. Astrophys. {\bf 447}, 31 (2006)
\bibitem{6} R. R. Caldwell, W Knowp, L Parker and D A T Vanzella, Phys. Rev. D {\bf 73}, 023513 (2006)
\bibitem{7} S. M. Carroll, W H Press and E L Turner, Ann. Rev. Astron. Astrophys. {\bf 30}, 499 (1992)
\bibitem{8} E. J. Copeland, M Sami and S Tsujikawa, Int. J. Mod. Phys. D  {\bf 15}, 1753 (2006)
\bibitem{9} A. Freedman et al., Astrophys. J. {\bf 553}, 47 (2001)
\bibitem{10} G. K, Goswami, A. K. Yadav and M. Mishra, Int. J. Theor. Phys. {\bf 54}, 315 (2015)
\bibitem{11} \O{}. Gr\O{}n and S. Hervik, $\it{Einstien's~ General~ Theory~ of~ Relativity:~ With~ Modern~
Application~ in~ Cosmology}$ ( Springer,  2007).
\bibitem{12 } G. Hinshaw et al., Astrophys. J. Suppl. {\bf 208}, 19 (2013)
\bibitem{13} T. Koivisto and D. F. Mota,`` Anisotropic dark energy: dynamics of the background and perturbations'', arXiv: 0801.3676.
\bibitem{14} E. Komastu et al., Astrophys. J. Suppl. Ser. {\bf 180}, 330 (2009)
\bibitem{15} S. Kumar and A. K. Yadav, Mod. Phys. Lett. A {\bf 26}, 647 (2011)
\bibitem{16} C. W. Misner, Astrophys. J. {\bf 151}, 431 (1968)
\bibitem{17} T. Koivisto and D. F. Mota, Astrophys. J. {\bf 679}, 1 (2008)
\bibitem{18} S. Perlmutter et al., Astrophys. J. {\bf 517}, 565 (1999)
\bibitem{19} A. Pradhan, Res. Astron. Astrophys. {\bf 13}, 139 (2013)
\bibitem{20} A. G. Riess et al., Astron. J. {\bf 116}, 1009 (1998)
\bibitem{21} A. G. Riess et al., Astron. J. {\bf 607}, 665 (2004)
\bibitem{22} B. Saha and A. K. Yadav, Astrophys. Space Sc. {\bf 341}, 651 (2012)
\bibitem{23} M. R. Setare and E. N. Saridakis, Phys. Lett. B {\bf 668}, 177 (2008)
\bibitem{24} M. R. Setare and E. N. Saridakis, JCAP {\bf 0903}, 002 (2009)
\bibitem{25} D. N. Spergel et al., Astrophys. J. Suppl. Ser. {\bf 148}, 175 (2003)
\bibitem{26} N. Suzuki et al., Astrophys. J. {\bf 746}, 85 (2012)
\bibitem{27} A. K. Yadav and L. Yadav, Int. J. Theor. Phys. {\bf 50}, 218 (2011)
\bibitem{28} A. K. Yadav, F. Rahaman and S. Ray, Int. J. Theor. Phys. {\bf 50}, 871 (2011)
\bibitem{29} A. K. Yadav, F. Rahaman, S. Ray and G. K. Goswami, Euro. Phys. J. Plus {\bf 127}, 127 (2012)
\bibitem{30} A. K. Yadav, Astrophys. Space Sc. {\bf 335}, 565 (2012)
\bibitem{31} O. Heckmann and E. Schucking, $\it{ Gravitation: ~An~ Introduction~ to~ current~ research}$ ( Willey, New York,
1962).
\end{thebibliography}
\end{document}